\documentstyle[epsf,prl,aps,floats]{revtex}

\newcommand{\Usub}{{\mbox{\scriptsize K/Si-B}}}
\begin{document}
\draft

\wideabs{ 

\title{
Strongly Correlated Electrons on a Silicon Surface: Theory of a Mott Insulator}
\author{ C. Stephen Hellberg\cite{email}
and Steven C. Erwin\cite{now-in-berlin}}
\address{
Complex Systems Theory Branch, Naval Research Laboratory, Washington,
D.C. 20375}
\date{\today}
\maketitle
\begin{abstract}
We demonstrate theoretically that the electronic ground state of the
potassium-covered Si(111)-B surface is a Mott insulator, explicitly
contradicting band theory but in good agreement with recent
experiments. We determine the physical structure by standard
density-functional methods,
and obtain the electronic ground state by exact diagonalization of a
many-body Hamiltonian.  The many-body conductivity reveals a
Brinkman-Rice metal-insulator transition at a critical interaction
strength; the calculated interaction strength is well above this
critical value.
\end{abstract}
\pacs{
73.20.-r, 
71.30.+h, 
71.10.Fd, 
71.27.+a  
}

}

Transport behavior in crystalline materials is governed by the
excitation spectrum: insulators have a finite gap to excitations while
metals have zero-energy excitations.  Band theory accurately describes
this distinction in most materials: systems with only filled or empty
bands are insulating while systems with partially occupied bands are
metallic.  However, the band description may break down under
circumstances when, roughly speaking, the energy cost for forming an
extended state exceeds the cost for forming a localized state. The
resulting ground state, which arises from electron-electron
interactions that band theory cannot describe, is known as a Mott
insulator
\cite{brinkman70a,gebhard97,imada98a}.

Surfaces provide a potentially fertile environment for Mott
insulators\cite{carpinelli96a}.  Electrons occupying surface states
may localize more readily than in the bulk, due to two significant
effects: (1) Atoms at surfaces have lower coordination than in the
bulk, raising the energetic cost for electron hopping. (2) Surfaces
often undergo reconstructions, yielding much larger inter-orbital
spacings than in the bulk.  These effects combine to make surfaces
natural systems to look for Mott insulating behavior.  In a recent
series of experiments, Weitering {\it et al.}
\cite{weitering97a,weitering93a} used photoemission and inverse
photoemission to demonstrate that the
K/Si(111)-$(\sqrt{3}\times\sqrt{3})$-B surface has a gap at the Fermi
level. Since this system has an odd number of electrons per unit cell
{\it it must be metallic in a band description}, clearly contradicting
the photoemission data.  On this basis, Weitering {\it et al.}\
hypothesized that this system (hereafter K/Si-B) is a Mott insulator.

In this Letter we explicitly demonstrate, by exact solution of the
appropriate many-body Hamiltonian, that the electronic ground state of
K/Si-B is indeed a Mott insulator.  The calculation is in three
parts. First, we use standard density-functional methods to determine
the geometrical and electronic structure of this surface within the
local-density approximation (LDA). Second, we map the relevant
electronic states onto a many-body Hamiltonian, which we then solve on
a periodic cluster using exact diagonalization techniques. Third, we
use the resulting many-body ground state to compute the zero-frequency
conductivity or Drude weight, $D$, and then show that in the infinite
system $D\rightarrow 0$, that is, a metal-insulator transition occurs
in the thermodynamic limit.

\begin{figure}[tbp]
\epsfxsize=6.0cm\centerline{\epsffile{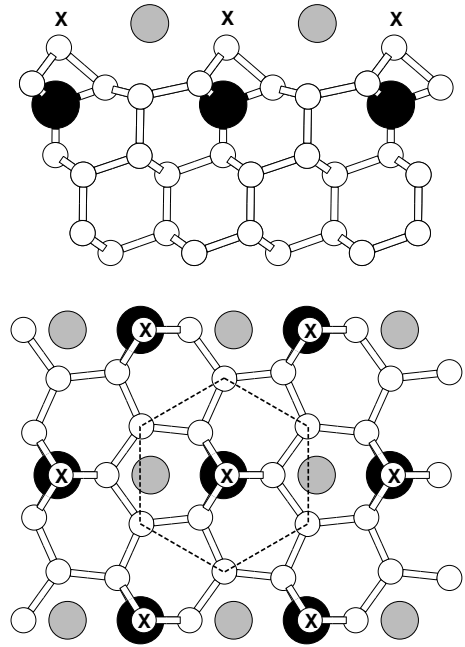}}
\vspace{.1in}
\caption{Side and top views of the fully relaxed structure of 
K/Si(111)-$(\sqrt{3}\times\sqrt{3})$-B.  K atoms are gray, Si atoms
are white, and B atoms are black.  The hexagon outlines the unit cell,
and {\bf X}'s denote Si dangling-bond orbitals. }
\label{fig:structure}
\end{figure}

Boron induces a well-known $\sqrt{3}\times\sqrt{3}$ reconstruction of
the clean Si(111) surface \cite{lyo89a}.  Boron substitutes for every
third Si atom in the second subsurface layer, and the displaced Si
assumes an adatom position above the boron (see Fig.\
\ref{fig:structure}).  The electron in the Si-adatom dangling bond is
transferred subsurface, enabling the B atom to participate in four
covalent bonds.  By this mechanism, the surface forms a conventional
band insulator, leaving each Si adatom with an empty orbital extending
away from the surface.  These orbitals form a triangular lattice on
the surface.  In the experiments of Weitering {\it et al.}, K was then
deposited onto this insulating substrate until the saturation coverage
was reached.

To determine the equilibrium structure of K/Si-B, we have performed
extensive LDA calculations. The calculations used a slab geometry with
three double layers of Si, terminated by H, and a vacuum region
equivalent to three double layers of Si.  Total energies and forces
were calculated using Hamann and Troullier-Martins pseudopotentials,
and a plane-wave basis with a kinetic-energy cutoff of 20 Ry, as
implemented in the {\sc fhi96md} code \cite{bockstedte97a}.  Four
k-points were used for Brillouin-zone integrations. Full structural
relaxation was performed on all atoms, except those in the bottommost
double layer, until the rms force was less than 0.05 eV/\AA.  We began
by first fully relaxing the surface without K present, and then
proceeded to determine the equilibrium coverage and geometry of the
K-saturated surface.

Experimentally, coverage is monitored via the electron work function:
at the saturation coverage, the low-temperature work function reaches
a minimum \cite{weitering93a}.  The absolute K coverage is not known
from experiment, so it must be determined theoretically. We calculated
the work function for the lowest-energy arrangement of adsorbates at
coverages of 1/6, 1/3, 2/3, and 1 monolayer (ML), and find a minimum
at 1/3 ML, in agreement with the conclusions of Weitering {\it et al.}
At all coverages, the experimental photoemission spectra show that the
Si adatom backbond state persists upon K deposition, suggesting the K
adsorbates do not break the Si adatom bonds\cite{weitering93a}.  We
therefore assume that at these coverages the adsorbates do not destroy
the underlying reconstruction.  The resulting minimum energy
configuration at the saturation coverage of 1/3 ML is shown in Fig.\
\ref{fig:structure}. The K adsorbates are in the $H_3$ hollow site,
with the Si adatom slightly shifted from its position with no K
present.  A metastable state with the K adsorbate in the $T_4$ hollow
site has an energy 0.1 eV higher per unit cell.

\begin{figure}[tb]
\epsfxsize=6.0cm\centerline{\epsffile{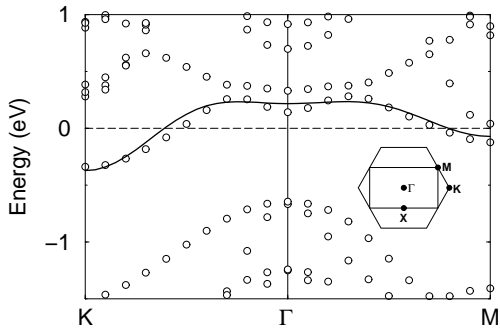}}
\caption{LDA band structure of K/Si-B at the K saturation coverage of
1/3 ML.  The Si-related band crossing the Fermi level is half occupied
by the K 4$s$ electron. The solid curve is the optimized fit of the
mean-field Hubbard dispersion, Eq.\ (\ref{eq:mf}), to the LDA
eigenvalues.  The inset shows the Brillouin zone of the fundamental
unit cell and of the doubled supercell, used to compute $U_\Usub$.  }
\label{fig:band}
\end{figure}

At coverages below 2/3 ML, one expects the K 4$s$ electrons to
partially occupy the surface state arising from the empty Si-adatom
orbitals.  At 1/3 ML there is one K per Si orbital, so in the band
description the single surface band is half occupied; this simple
picture is confirmed by the calculated LDA band structure shown in
Fig.\ \ref{fig:band}.  Clearly this system must be metallic within
band theory.  To investigate the importance of electronic interactions
not included in band theory, we derive a single-band Hubbard model for
the half-filled surface state.  The Hamiltonian is
\begin{equation}
 H = \sum_{i j \sigma} t_{ij}
  c_{ i \sigma}^\dagger c_{ j \sigma}^{\mbox{}}
+ U \sum_i n_{i\uparrow} n_{i\downarrow},
\label{eq:hubbard}
\end{equation}
where $ c_{ i \sigma}^\dagger$ creates an electron with spin $\sigma$ on
site $i$, and $ n_{i\sigma} = c_{i\sigma}^{\dagger}c_{i\sigma}^{\mbox{}}$
is the number operator.
The sites correspond to the empty Si orbitals that the K
electrons are doping.
The amplitude for hopping from orbital $i$ to orbital $j$ is given by
$t_{ij}$, and $t_{ij} = t_{ji}$.
There is a Coulomb energy cost of $U$ to occupy an orbital with
two electrons.

Our approach to determining the parameters of the Hubbard model is
similar to other first-principles approaches
\cite{pankratov93a,hybertsen89a}. We first solve the Hubbard model in the
mean-field (MF) approximation, and then require that the resulting
single-particle energies optimally reproduce the corresponding LDA
spectrum (which is also a mean-field theory) throughout the zone. In
the MF approximation, the up electrons move in the average potential
generated by the down electrons (and vice-versa), so the MF Hubbard
Hamiltonian for the up electrons becomes
\begin{equation}
 H_\uparrow^{{\rm MF}} = \sum_{i j} t_{ij}
  c_{ i \uparrow}^\dagger c_{ j \uparrow}^{\mbox{}}
+ U \sum_i n_{i\uparrow} \langle n_{i\downarrow} \rangle ,
\label{eq:mf}
\end{equation}
where $ \langle n_{i\downarrow} \rangle $ is the average density of
the down electrons on site $i$.  We assume a paramagnetic state in
both the LDA and the MF solution to the Hubbard model.

To determine the hopping amplitudes, $t_{ij}$, we fit the
single-particle eigenvalues in the MF Hubbard solution to the LDA
eigenvalues of the surface band at 100 special $k$-points.  Note that
although the K adsorbates break the three-fold rotational symmetry of
the substrate, the LDA band structure remains nearly isotropic, and so
we assume isotropic hopping.  We allow hopping between
nearest-neighboring and second-nearest-neighboring sites, with
amplitudes $t_1$ and $t_2$ respectively---Third-nearest-neighbor hopping
was found to be insignificant.  
Thus the dispersion is given by
\begin{eqnarray}
\varepsilon({\bf k})
& = & 2 t_1
\left[ \cos(k_y) + 2 \cos(k_x \sqrt{3}/2) \cos(k_y/2) \right]
\nonumber
\\
& + & 2 t_2
\left[ \cos(k_x \sqrt{3}) + 2 \cos(k_x \sqrt{3}/2) \cos(3 k_y/2) \right] .
\label{eq:dispersion}
\end{eqnarray}
The optimized amplitudes are $t_1 = 66$ meV and $t_2 = -24$ meV.  A plot
of the fit and the LDA eigenvalues along high symmetry directions is
shown in Fig.\ \ref{fig:band}; the fit is very good, with a rms error
of 42 meV.

\begin{figure}[tb]
\epsfxsize=7.5cm\centerline{\epsffile{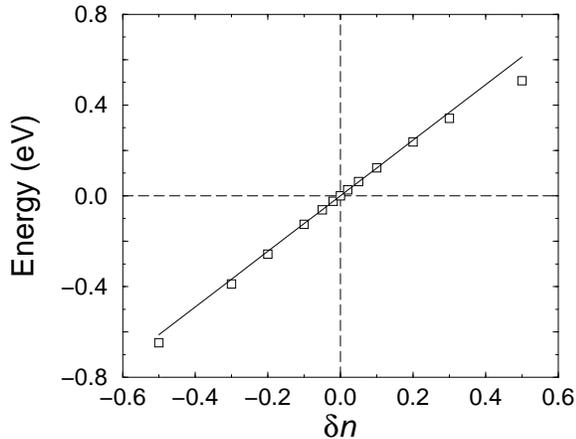}}
\caption{
The single-particle energy $\varepsilon_+$ at the edge of the
Brillouin zone (the X point), as a function of the electron transfer
$\delta n$.  The squares are LDA supercell results; the solid line is
the best fit of Eq.\ (\ref{eq:U}) to the LDA results for $|\delta n|
\leq 0.3$.  }
\label{fig:potential}
\end{figure}

To determine the intra-orbital Coulomb repulsion $U$, we subject the
system to a density fluctuation by moving charge from one Si orbital
to another.  The optimal interaction parameter, $U_\Usub$, is then
determined by requiring the LDA solution and the MF Hubbard
calculation to respond identically.  In order to maintain overall
charge neutrality, a supercell calculation with two Si orbitals is
required.  At the edge of the Brillouin zone, the paramagnetic
single-particle eigenvalues for a shift of $\delta n$ of an electron
from one orbital to another take the simple form
\begin{equation}
\varepsilon_\pm = \pm  U \, \delta n/2 ,
\label{eq:U}
\end{equation}
up to an overall constant which we take to be zero.  The fit of
$\varepsilon_+$ to the LDA eigenvalues is shown in Fig.\
\ref{fig:potential}.  For small charge shifts, the LDA eigenvalues are
nearly linear, and we obtain $U_\Usub=1.23$ eV.  For larger charge
shifts, additional bands in the LDA calculation enter that are not
present in the single-band Hubbard model, and the LDA eigenvalues drop
very slightly below the MF Hubbard eigenvalues.  As a check of the
reliability of this approach, we also determined $U_\Usub$ by fitting
the change in the total kinetic energy in the MF Hubbard solution to
the LDA.  We find $U_\Usub \approx 1.2$ eV, consistent with the above
result.

Having determined the parameters of the Hubbard model describing
K/Si-B, we now solve this model exactly (using a periodic 16-site
cluster) to obtain the many-body electronic ground state.  The number
of states in the Hilbert space grows exponentially with the number of
sites in the cluster: in the Hubbard model, each site can have zero,
one (either up or down), or two electrons, so the number of basis
states in the space of an $N$-site system is $4^N$.  The symmetries of
the Hamiltonian make the matrix block diagonal, but with 16 sites the
size of the largest block is still more than $10^7\times 10^7$.
Conventional algorithms obviously cannot diagonalize matrices this
large, and so we use the Lanczos algorithm to determine the exact
ground state\cite{cullum85a}.  Storing the Hamiltonian and three basis
vectors in memory on an IBM SP2 required 64 nodes with 1 Gb memory
each; the computation required about 600 CPU hours.

\begin{figure}[tb]
\epsfxsize=7.5cm\centerline{\epsffile{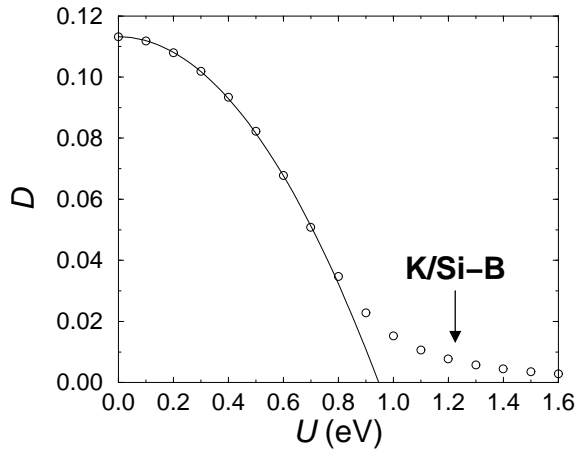}}
\caption{The Drude weight, $D$,
as a function of the interaction parameter $U$.  The circles are exact
results from the Hubbard model on a 16-site cluster, and the curve is
a fit of the Hubbard results (for $U \leq 0.7$ eV) to the infinite
limit given by Eq.\ (\ref{eq:drude}).  }
\label{fig:drude}
\end{figure}

To distinguish quantitatively between metallic and insulating
behavior, we calculate the zero-frequency conductivity or Drude
weight, $D$, of the many-body ground state.  The Drude weight provides
a definitive way to distinguish metals from insulators irrespective of
the applicability of band theory: in the thermodynamic limit, metals
have non-zero Drude weight, while for insulators $D=0$.  Kohn first
showed that $D$ may be calculated from the variation of the
ground-state energy with respect to an applied vector potential
\cite{gebhard97,imada98a,kohn64a},
\begin{equation}
D = \partial^2E / \partial \phi^2.
\end{equation}
We use this technique to determine the Drude weight of our Hubbard
cluster as a function of the interaction parameter $U$.  The results
are shown in Fig.\ \ref{fig:drude}.  At small $U$, the system is in
the metallic regime and so $D$ is large, and it decreases
monotonically with increasing $U$.  $D$ never becomes zero, because
only an infinite system can undergo a true metal-insulator transition.
Instead of a critical interaction strength, our 16-site cluster shows
a transition {\it region} in the vicinity of the physically
interesting interaction strengths, near $U=1$ eV.

In general, there may be level crossings in the ground state 
as a function of $U$.
For our hopping amplitudes, this is not the case---The 
ground state evolves adiabatically with increasing $U$,
and the Drude weight in Fig.\ \ref{fig:drude} is a continuous function.
We see no evidence of the intermediate (semi-metallic) phase 
that was found in recent slave-boson studies of the Hubbard model 
on a triangular lattice with $t_2=0$ \cite{gazza94a}.

To show that K/Si-B lies on the insulating side of a Mott-insulator
transition, we need the critical interaction strength for the
transition in order to compare with our calculated interaction
strength, $U_\Usub$.  We can obtain this value by extending our
calculated $D(U)$ from the exact finite-cluster result to the
thermodynamic infinite limit, using the form derived with the
Gutzwiller approximation\cite{brinkman70a} for the infinite system,
\begin{equation}
D_\infty(U) \propto
\left\{ \begin{array}{ll}
1 - (U/U_c)^2, & ~ U < U_c \\
0,             & ~ U > U_c
\end{array} \right.
\label{eq:drude}
\end{equation}
where $U_c$ is the critical interaction strength for the
metal-insulator transition.  This functional form fits our 16-site
results at small and intermediate values of $U$ extremely well, as
shown in Fig.\ \ref{fig:drude}.  From this fit, we estimate the
critical interaction strength for the Mott metal-insulator transition
in the infinite system occurs at $U_c = 0.95\pm0.02$ eV.  Our
calculated value for the physical system, $U_\Usub=1.23$ eV, is well
above this critical value, and establishes that K/Si-B is indeed a
Mott insulator.

In the Mott-insulating regime, the localized electrons will interact
via an antiferromagnetic Heisenberg exchange coupling with strength
$J_{ij} = 4 t_{ij}^2/U$ \cite{auerbach94}.  The Heisenberg model on a
triangular lattice is frustrated.  The nearest-neighbor model has
strong three-sublattice correlations, but whether the correlations are
long-ranged is controversial \cite{huse88a}. In the Hubbard model for
K/Si-B, the second-neighbor hopping is substantial, so the
second-neighbor antiferromagnetic Heisenberg coupling in the
Mott-insulating limit will be significant.  This coupling frustrates
the three-sublattice correlations in the nearest-neighbor model, so it
is unlikely that three-sublattice order is established.  The ground
state will likely either establish some other type of collinear order
\cite{jolicoeur90a} or enter a quantum-disordered regime with no
long-range order
\cite{gebhard97}.

To conclude, we have shown that the many-body electronic ground state
of the K/Si-B surface is a Mott insulator. Specifically, we have first
determined the surface coverage and morphology of K adsorbed on
Si(111)-B using first-principles total-energy methods.  We then mapped
the relevant electronic degrees of freedom onto a Hubbard model, which
we solved with exact diagonalization.  By calculating the Drude weight
of the model, we have demonstrated that for the physical parameters of
K/Si-B, the model has a Mott insulating ground state.

We thank H. H. Weitering, E.~J. Mele, M.~J. Rozenberg, and D.~W. Hess for
enlightening discussions.  This work was supported by the National
Research Council and was funded by ONR\@.  Computational work was
supported by a grant of HPC time from the DoD Major Shared Resource
Center \mbox{ASCWP}\@.

\end{document}